\documentclass[runningheads]{llncs}
\usepackage{graphicx}
\usepackage{booktabs}
\usepackage{graphicx,calc}
\newlength\myheight
\newlength\mydepth
\settototalheight\myheight{Xygp}
\usepackage{comment}
\usepackage{wrapfig}
\usepackage{subcaption}
\usepackage[font=small]{caption, subcaption}
\usepackage{todonotes}

\usepackage[hyphens]{url}  
\usepackage[section]{placeins}

\begin{document}
\title{ALONE: A Dataset for Toxic Behavior among Adolescents on Twitter}
%
%

\author{
\textbf{Thilini Wijesiriwardene}\textsuperscript{\rm 1}\textsuperscript{\rm *}
\textbf{Hale Inan}\textsuperscript{\rm 4}\textsuperscript{\rm *}
\textbf{Ugur Kursuncu}\textsuperscript{\rm 1} \\
\textbf{Manas Gaur}\textsuperscript{\rm 1}
\textbf{Valerie L. Shalin}\textsuperscript{\rm 2}
\textbf{Krishnaprasad Thirunarayan}\textsuperscript{\rm 3}
\textbf{Amit Sheth}\textsuperscript{\rm 1}
\textbf{I. Budak Arpinar}\textsuperscript{\rm 4}\\ 
}

\authorrunning{Wijesiriwardene, Inan and Kursuncu et al.}
%
\institute{AI Institute, University of South Carolina\\ 
\email{thilini@sc.edu, kursuncu@mailbox.sc.edu, mgaur@email.sc.edu, amit@sc.edu} 
\and
Department of Psychology, Wright State University\\
\email{valerie.shalin@wright.edu}\\
\and
Department of Computer Science and Engineering, Wright State University\\
\email{t.k.prasad@wright.edu}
\and
Department of Computer Science, University of Georgia\\
\email{hale.inan25@uga.edu, budak@cs.uga.edu}\\
\textsuperscript{\rm *}\textit{Equally contributed.}
}

%
\maketitle              
\begin{abstract}
\vspace{-4mm}
The convenience of social media has also enabled its misuse, potentially resulting  in toxic behavior. Nearly 66\% of internet users have observed online harassment, and 41\% claim personal experience, with 18\% facing severe forms of online harassment. This toxic communication has a significant impact on the well-being of young individuals, affecting  mental health and, in some cases, resulting in suicide. These  communications exhibit complex linguistic and contextual characteristics, making recognition of such narratives challenging. In this paper, we provide a multimodal dataset of toxic social media interactions between confirmed high school students, called ALONE (AdoLescents ON twittEr), along with descriptive explanation. Each instance of interaction includes tweets, images, emoji and related metadata. Our observations show that individual tweets do not provide sufficient evidence for toxic behavior, and meaningful use of context in interactions can enable highlighting or exonerating tweets with purported toxicity.

\keywords{Toxicity \and Harassment \and Social Media \and Resource \and Dataset}
\end{abstract}

\section{Introduction}
\label{sec:intro}
\vspace{-1mm}
The language of social media is a socio-cultural product, reflecting issues of relevance to the sample population and evolving norms in the exchange of coarse language and acceptable sarcasm, employing toxic, questionable language, and sometimes constituting actual harassment. According to a  2017 Pew Research Center survey, 41\% of U.S. adults claim to have experienced some  type of online harassment, offensive name-calling, purposeful embarrassment, physical threats, harassment over a sustained period of time, sexual harassment or stalking\footnote{\url{https://www.pewresearch.org/fact-tank/2017/07/11/key-takeaways-online-harassment/}}.

Toxic behavior is  prevalent among adolescents, sometimes leading to aggression \cite{liu2013understanding,lowry1998weapon}. Adolescents exemplify a population that is particularly vulnerable to disturbing social media interactions\footnote{\url{https://www.cim.co.uk/newsroom/release-half-of-teens-exposed-to-harmful-social-media/}} \cite{unicef2018everyday}, and this behavior is observable in a network of high school students \cite{brener1999recent}. Further, a toxic online environment may cause mental health problems for this population\footnote{\url{https://www.cnn.com/2016/12/14/health/teen-suicide-cyberbullying-continues-trnd/index.html}} \cite{arseneault2010bullying,rivers2009observing,kumpulainen2001psychiatric,viner2019roles}. While a victim may experience a negative reaction from a toxic environment of offensive language, this differs from \emph{targeted} toxicity which is usually directed whose content collected and confirmed with a unique method towards one individual. The analysis of single tweets or individual users is potentially  misleading as the context of interactions between the two people (e.g., source and target) dictates the determination of toxicity. In other words, two individuals who are friends may use coarse keywords or language that is seemingly toxic, but it may be sarcastic, exonerating them from toxicity. 

In this paper, we provide a dataset and its details, specific to toxic behavior in social media communications. This dataset has two particular contributions: (i) the population is \emph{high school students} whose content was collected and confirmed with a unique method, and (ii) it was designed based on the \emph{interactions} between participants. The detection of true toxic behavior against a persisting background of coarse language poses a challenging task.  Moreover, the scope of the original crawl has great bearing on the prevalence of toxicity features and the criteria for toxic behavior itself.  To address these issues, we have assembled a social media corpus from Twitter for a sample of midwestern American High School Students.  We assert a dyadic, directed interaction, between a source and a target. Existing related datasets (see Related Work section) focus mainly on the user or tweet level for the task of detecting toxic content. Such datasets fail to  capture adequately the fundamental and contextual nuances in the language of these conversations. Thus, our corpus preserves and aggregates the social media interaction history between participants. This enables the determination of existing friendship and hence possible sarcasm. Because individuals can communicate with multiple partners, we have the potential of detecting unique toxic person-victim pairings that would be otherwise undetectable in the raw original crawl.

Each entry in our dataset consists of 12 fields: \textit{Interaction Id, Count, Source User Id, Target User Id, Emoji, Emoji Keywords, Tweets, Image Keywords, created at, favorite count, in reply to screenname} and \textit{label} where the \textit{Tweets} field contains an aggregation of the tweets between a specific pair of source and target. For preliminary analysis, we define a single dimension of \emph{toxic language}, pegged at one end by benign content and the other by harassment. This dimension can be partitioned into several, partially overlapping classes, determined  by a decision rule. We have identified and experimented with three levels of toxic interactions between source and target: \textit{Toxic (T), Non-Toxic (N), or Unclear (U)}. However, the boundaries between levels are discretionary, accommodating construct definitions that are, at best, debatable.  

 We  include examples  across the continuum of toxic language, with sufficient context to determine the nature of toxicity. We detect true toxicity on Twitter by analyzing interactions among a collection of tweets, in contrast with prior approaches where the main focus is performing user or tweet level analysis. Further, we assert that detecting a user as a toxic person with respect to one victim does not provide evidence of being a universal toxic person because they can be friendly to a majority of others.

\section{Related Work}
\label{sec:RW}
\vspace{-2mm}
We reviewed prior work for the variety of overlapping constructs related to toxic exchanges.  The social media literature related to toxic behavior lacks crisp distinctions between: offensive language \cite{ICWSM1817909,jay2008pragmatics,razavi2010offensive}, hate speech \cite{ICWSM1817909,davidson2017automated,badjatiya2017deep,warner2012detecting}, abusive language \cite{ICWSM1817909,papegnies2017detection,nobata2016abusive} and cyberbullying \cite{chatzakou2017mean,hosseinmardi2015analyzing,dinakar2011modeling}. For example, the following definition of offensive language substantially overlaps with the subsequent definition of hate speech. {According to} \cite{ICWSM1817909}, \emph{offensive language is profanity, strongly impolite, rude or vulgar language expressed with fighting or hurtful words in order to insult a targeted individual or group. Hate speech is language used to express hatred towards a targeted individual or group, or is intended to be derogatory, to humiliate, or to insult the members of a group, on the basis of attributes such as race, religion, ethnic origin, sexual orientation, disability, or gender.}  \cite{salminen2018anatomy} classifies swearing, aggressive comments, or mentioning the past political or ethnic conflicts in a non-constructive and harmful way as hateful: \textit{@user\_name nope you just a stupid hoe who wouldn't know their place} \raisebox{-\mydepth}{\includegraphics[height=\myheight]{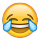}}\raisebox{-\mydepth}{\includegraphics[height=\myheight]{figures/tears_of_joy.pdf}} comprises both offensive and hate speech. Specifically, the challenge lies in operationalizing the contextual differences between offensiveness, hate speech and harassment. 
As the existing work on offensive content, harassment and hate speech fails to take into account the nature of the relationship between participants, we focus our attention on the context-aware analyses of  targeted exchanges. 

\textbf{Offensive}— \cite{badjatiya2017deep} annotated 16K tweets from \cite{waseem2016hateful} with the labels, racist, sexist or neither. 3383 and 1972 tweets were sexist and racist respectively, and others were labeled as neither. In \cite{nobata2016abusive}, their aim was to detect abusive language on online user comments posted on Yahoo. 56,280 comments were labeled as ‘Abusive’ and 895,546 comments as ‘Clean’.

\textbf{Hate Speech}—\cite{silva2016analyzing} developed a dataset to identify the main targets of online hate speech including the nine categories such as race, behavior, physical, sexual orientation, class, gender, ethnicity, disability, religion, and other for non-classified hate targets. 178 most popular targets from Whisper and Twitter were manually labeled, unveiling new forms of online hate that can be harmful to people. \cite{davidson2017automated} focused on distinguishing hate speech from other forms of offensive language. They extracted 85.4 million tweets from 33,458 users, and randomly sampled 25K tweets containing words from a hate speech lexicon. Individual tweets were labeled as hate speech, offensive or neither.  \cite{sharma2018degree} presented an annotated corpus of tweets classified by different levels of hate to provide an ontological classification model to identify harmful speech. They randomly sampled 14,906 tweets and developed a supervised system used for detection of the class of harmful speech. In \cite{waseem2016hateful}, tweets were sampled from the 130K tweets, and in addition to “racism”, “sexism”, and “neither”, the label “both” was added. A character n-gram based approach provided better performance for hate speech detection.  \cite{waseem2016you} examined the influence of annotators' knowledge for hate speech on classification models, labeling individual tweets. Considering only cases of full agreement among amateur annotators, they found that amateur annotators can produce relatively good annotations as compared to expert annotators.  

\textbf{Harassment}—A number of researchers have attempted to identify dimensions or factors underpinning harassment. \cite{namie2009bully} drew on the model \cite{buss1961psychology} that conceptualized aggression on four dimensions: \emph{verbal, physical, direct-indirect,} and \emph{active-passive}. \cite{rezvan2020analyzing} analyses the linguistics aspects of harassment based on different harassment types. Consistent with our interest in interaction history between participants, cyberbullying emphasizes the repetitiveness of aggressive acts \cite{patchin2006bullies}. The harasser may target a victim over a period of time, or a group of harassers may target a victim about the same demeaning characteristic or incident. Apart from repetitiveness, the difference of power between the harasser and victim suggests cyberbullying. However, this work \cite{patchin2006bullies} is not computationally oriented. Golbeck \cite{golbeck2017large} introduced a large, human labeled corpus of online harassment data including 35,000 tweets with 5495 non-harassing and 29505 harassing examples. 

In contrast to this literature, our approach to the problem is to focus on interactions between participants to capture the context of the relationship rather than solely tweets or users. As online toxic behavior is a complex issue that involves different contexts and dimensions \cite{kursuncu2019modeling,kursuncu2018modeling,arpinar2016social}, 
tweet-level or user-level approaches do not adequately capture the context with important nuances due to the fluidity in the language. Our interaction-based dataset will enable researchers to uncover critical patterns for gaining a better understanding of toxic behavior on social media. Additionally, our dataset is unique in its focus on high school student demographic.

\section{Dataset}
\label{sec:dataset}
\vspace{-2mm}
\noindent For the dataset ALONE, we retrieved 469,786 tweets from our raw Twitter data, and used a harassment lexicon provided by \cite{rezvan2018quality} to filter tweets that are likely to contain toxic behavior, obtaining a collection of 688 interactions with aggregated 16,901 tweets.  

\subsection{Data Collection}
\label{sec:data-collect}
We focused on tweets as the source for our dataset because of its public access. Besides text, tweets can contain images, emoji and URLs as additional content. To create a ground truth dataset, we reviewed public lists of students, such as the list of National Merit Scholars published in newspapers, identifying 143 names of the attendees of a high school. Using the list of identified individuals, we searched Twitter for the profiles associated with these students using Twitter APIs. Then, with the guidance of our cognitive scientist co-author, we confirmed that the users that we retrieved were high school students, through their profiles and tweets conversing on their school mascot, clubs or faculty members. The 143 user profiles with their tweets constituted the seed corpus. 

\paragraph{\textbf{Dataset Expansion:}} As a typical network of high school students is  larger than 143 users, we expanded the network using the friend and follower relationships. We followed the following procedure: 
\begin{quote}
    \begin{itemize}
        \item Collect friends and followers lists for each seed profile.
        \item Exclude non-student accounts: We identified the accounts following each other considering them as candidate students, and removed accounts that are not both following and being followed by the accounts in the friends and followers lists of seed accounts (not common profiles). As the adults, such as teachers, would  notice  any toxic behavior, such as harassment, bullying or aggression, which may have consequences, students with potentially toxic behavior would avoid following their social media accounts \cite{sondergaard2012bullying,nilan2015youth} to sequester  social network behavior \cite{mishna2014students,nilan2015youth}. We obtained 8805 accounts that follow and are being followed by at least one seed account, as candidates for student accounts in the high school. We removed 80 accounts as  they were suspended or deleted or otherwise protected by account owners.
        \item Retain only the peer profiles that follow and are being followed by more than 10\% of the seed profiles, yielding 320 likely peers. To confirm the absence of false positives, 50 accounts out of the 320 likely peers were randomly selected and manually validated that all the 50 were confirmed student accounts. When tweets of the newly added 320 accounts were crawled, seven accounts were deleted or restricted. Hence, we removed them from the dataset, resulting in 456 accounts (143 seed and 313 added).
    \end{itemize}
\end{quote}

\noindent After we finalized the 456 accounts, tweets (up to 3200 if available) were collected for each, starting from the most recent (May 2018), along with their account metadata, using the Twitter API. 

\paragraph{\textbf{Interaction-based Dataset:}} As our toxic behavior construct requires interactions between participants, we pruned the tweet corpus to retain a dataset that consists of interactions. We define an interaction as a collection of tweets exchanged between the two participants (e.g., source and target) in one direction, and on Twitter, we consider mentions (including replies) and retweets as interactions. For instance, one user may mention another user in a tweet for harassing, bullying or insulting. Moreover, retweeting a harassing tweet potentially boosts popularity, which creates the role of bystander for the source, suggesting that the retweeting user (source) is actually supporting or helping the harasser (target). We have left retweet indicators (e.g., RT @username:) in the data. Further, some tweets are included in multiple interactions; hence, these communications are a part of a group communication that is not dyadic. For some instances, source and target are the same users, and we left these conversations in the dataset as they may be likely a part of group aggression.

We aggregated tweets that qualify as interactions between users, potentially reducing the false alarm rate of an analysis solely based on the presence of characteristics of offensive language \cite{badjatiya2019stereotypical}. This allows for the detection of a  particularly intriguing combination of positive and negative sentiment lexical items, suggestive of sarcasm, e.g., \textit{happy birthday @user\_name love you but hate your feet} \raisebox{-\mydepth}{\includegraphics[height=\myheight]{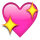}}\raisebox{-\mydepth}{\includegraphics[height=\myheight]{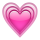}} \raisebox{-\mydepth}{\includegraphics[height=\myheight]{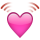}} \raisebox{-\mydepth}{\includegraphics[height=\myheight]{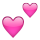}} \raisebox{-\mydepth}{\includegraphics[height=\myheight]{figures/heart3.pdf}} and \textit{Happy birthday ugly!!} \raisebox{-\mydepth}{\includegraphics[height=\myheight]{figures/heart4.pdf}}.
The presence of “Happy Birthday” or positive emoji (see above) alters the interpretation of content that would otherwise be regarded as potentially suggestive of toxic behavior and the phenomenon of conflicting valence \textit{exoneration} content, assuming that the toxic content is sarcastic, e.g., the source does not really believe the recipient has unattractive feet or is generally ugly. Moreover, contextual analysis reveals that some of these are not truly toxic. Prior tweets in an interaction provides exonerating context, by indicating the presence of friendship, thus correcting the false positive. Designing the dataset based on interactions captures the context of the relationship between the two user; thus, enabling one to employ computational techniques to retrieve meaningful information concerning true toxicity. 

A portion of the tweets does not include any interaction indicator, but they refer to a person indirectly without mentioning or writing the name with malicious intent, to avoid the authority figures. This is called \emph{Subtweeting}\footnote{\url{https://www.theguardian.com/technology/blog/2014/jul/23/subtweeting-what-is-it-and-how-to-do-it-well}}\footnote{\url{http://bolobhi.org/abuse-subtweeting-tweet-school-cyber-bullying/}} \cite{rafla2014adolescents,edwards2016tweet,crumback2017subtweet}. Adolescents have specifically developed such practice due to their own privacy concerns and parental intrusion. For each user, we aggregated the tweets that do not mention the target explicitly, and indicated the target as \emph{``None”}. 

Then, a harassment lexicon \cite{rezvan2018quality} was utilized to filter the interactions that potentially contain toxic content. For online harassment, source and target dyads can be considered as \emph{harasser-victim} or \emph{bystander-harasser}. Further, as capturing context to determine the toxicity in the content is critical, an interaction should include a sufficient number of tweets. Therefore, we set an empirical threshold for one interaction as having at least three tweets, to capture context.

We have fully \emph{de-identified} the interactions by replacing; (i) Twitter usernames and mentions in tweets with a numeric user id, (ii) URLs with the token of $<url>$, and (iii) person names with the token of $<name>$. We have also included the following metadata for each tweet in the interactions: timestamp, favorite counts, and the de-identified user id of the replied user (if the tweets is a reply). Thus, researchers will have the ability to study a variety of aspects of this problem such as time series analysis. The finalized dataset includes 688 interactions with 16,901 tweets. The fields in an instance are as follows: Interaction Id, Count, Source User Id, Target User Id, Label, Emoji, Emoji Keywords, Tweets, Image Keywords, Timestamp, in reply to and favorite count. ``Count” field holds information for the number of tweets in an interaction. ``Source and Target User Id” fields hold numeric identification \emph{(after de-identification)} information. A ``Label” field holds the assigned label (T,N,U) for the interaction. While the ``Emoji” field holds the emoji being used in the tweets, ``Emoji Keywords” field provides the keywords that explain the meaning of the emoji, retrieved from EmojiNet \cite{wijeratne2017emojinet}. The ``Tweets” field has the tweets, and the following fields holds the metadata for each tweet: (i)  Timestamp: time information of a tweet, (ii) in reply to: (non-real) user id of the target if the tweet is a reply, (iii) favorite count: number of favorites. See Table \ref{tab:examples_dataset} for example interactions from the dataset with four fields.

\begin{table}[hbt!]
\footnotesize
\begin{center}
\begin{tabular}{p{1cm}p{11cm}}
    \toprule[2.5pt]
     \textbf{Label} & \textbf{Tweets} \\ \midrule[1pt]
     T & if you gon say n…. this much, the LEAST you could do is hit the tanning bed $<url>$ ***   you're f…... the most hideous and racist piece of s... ***  YOU ARE LITERALLY F…... RACIST SHUT THE F… UP ***  yeah \raisebox{-\mydepth}{\includegraphics[height=\myheight]{figures/tears_of_joy.pdf}}\raisebox{-\mydepth}{\includegraphics[height=\myheight]{figures/tears_of_joy.pdf}} you're not racist at all !!!!!!!! *** are you in f…... politics no, you're like 17 s... the f... up and stop putting your \"facts\" on…\\
     \midrule[1pt]
     T & ight f... you again ***  nah f... all  of you frfr bunch of f…... f…... ***  f... you $<url>$ you have no room to be talking s... shut your bum a.. up frfr \raisebox{-\mydepth}{\includegraphics[height=\myheight]{figures/tears_of_joy.pdf}} **  you're halarious, f... you and everyone that favorited that and retweeted that \\ 
     \midrule[1pt]
     N & ``Kix is the handjob of cereals''- John Doe \raisebox{-\mydepth}{\includegraphics[height=\myheight]{figures/tears_of_joy.pdf}}\raisebox{-\mydepth}{\includegraphics[height=\myheight]{figures/tears_of_joy.pdf}} $<image url>$ ***  Explain to that i….. that doing it spreads the word and the chance of someone donating XD fedora wearing as… *** get the f... off my twitter b…. BOI ***  guys follow bc he's an i…. and forgot his password. \\
     \midrule[1pt]
      U & This tweet was dumb I agree with u this time *** hahaha I'm so dumb ***  that's my mom f.... ***  boob *** never seen a bigger lie on the Internet then this one right here
     \\ \bottomrule[2pt]
\end{tabular}
\end{center}
\caption{Examples from the dataset with labels Toxic (T), Non-Toxic (N) and Unclear (U). The expletives were replaced with the first letter followed by as many \emph{dots} as there are remaining letters.}
\label{tab:examples_dataset}
\vspace{-10mm}
\end{table}
\vspace{-1.1em}

\paragraph{\textbf{Multimodality:}} As it will be described in Section Descriptive Statistics, different modalities of data, such as text, image, emoji, appear  in Toxic and Non-Toxic interactions with different proportions. Therefore, we provided explanations of potentially valuable emoji and images. Each image name was created by combining ``source user id", ``target user id", and ``tweet number" in an interaction that each image pertains to. For example: the image 0023.0230.5.jpg is from a tweet between ``user 0023" and ``user 0230" and the 5th tweet in their interaction. We processed these images utilizing a state-of-the-art image recognition tool, ResNet\footnote{\url{https://github.com/onnx/models/tree/master/vision/classification/resnet}} \cite{he2016deep}, providing the objects recognized in images with their probabilities (top-5 accuracy= 0.921). We kept the top 20 (empirically set) recognized object names. For example, an image has the following set of recognized objects: ``television", ``cash machine", ``screen", ``monitor", ``neck brace", ``toyshop", ``medicine chest", ``library", ``home theater", ``wardrobe", ``scoreboard", ``moving van",``entertainment center", ``barbershop", ``desk", ``web site". We utilized EmojiNet \footnote{\url{http://wiki.aiisc.ai/index.php/EmojiNet}} \cite{wijeratne2017emojinet} to retrieve the meanings of the emoji in the interactions, and provided in the dataset. For instance, for the emoji \raisebox{-\mydepth}{\includegraphics[height=\myheight]{figures/tears_of_joy.pdf}}, EmojiNet provides the following set of keywords: ``face", ``tear", ``joy", ``laugh", ``happy", ``cute", ``funny", ``joyful", ``hilarious", ``teary", ``laughing", ``person", ``smiley", ``lol", ``emoji", ``wtf", ``cry", ``crying", ``tears", ``lmao". Specifically, the significant difference in the use of image, video and emoji between the content of Toxic and Non-Toxic interactions, suggests that the contribution of multimodal elements would likely be critical.

\begin{wraptable}{R}{0.41\textwidth}
\vspace{-2em}
\begin{tabular}{p{2.4cm}p{2.4cm}} \toprule[1.5pt]
    \textbf{Three Label} & \textbf{Two Label} \\ \midrule
    0.63 & 0.65 \\
    \bottomrule[1.5pt]
\end{tabular}
\vspace{-2mm}
\caption{\footnotesize For \textbf{three} and \textbf{two} labels, agreement scores between the three annotators using \textbf{Krippendorff’s alpha}.} 
\label{tab:kannotation}

\vspace{1.5em}

\begin{tabular}{p{1.8cm}p{1.5cm}p{1.5cm}} \toprule[1pt]
    \textbf{Kappa} & \textbf{A} & \textbf{B} \\ \midrule
          \textbf{B} & 0.77 & - \\
          \textbf{C} & 0.52 & 0.62 \\ \midrule[1pt]
\end{tabular}
\vspace{-3mm}
\caption{\footnotesize Pairwise agreement for the \textbf{three} label scheme, agreement scores between the three annotators (A,B,C) using Cohen Kappa} 
\label{tab:cohenannotation1}
\vspace{-1em}
\vspace{-4mm}
\end{wraptable}

\paragraph{\textbf{Privacy and Ethics Disclosure:}} We use only public Twitter data, and our study does not involve any direct interaction with any individuals or their personally identifiable private data. This study was reviewed by the host institution's IRB and received an exemption determination. As noted above, we follow standard practices for anonymization during data collection and processing by removing any identifiable information including names, usernames, URLs. We do not provide any Twitter user or tweet id, or geolocation information. 
Due to privacy concerns and terms of use by Twitter, we make this dataset available upon request to the authors, and researchers will be required to sign an agreement to use it only for research purposes and without public dissemination.   

\begin{wraptable}{R}{0.41\textwidth}
\vspace{-2em}
\begin{tabular}{p{1.8cm}p{1.5cm}p{1.5cm}} \toprule[1pt]
    \textbf{Kappa} & \textbf{A} & \textbf{B} \\ \midrule
          \textbf{B} & 0.82 & - \\
          \textbf{C} & 0.49 & 0.63 \\
    \bottomrule[1.5pt] \\
\end{tabular}
\vspace{-7mm}
\caption{\footnotesize Pairwise agreement for \textbf{two} labels, agreement scores between the three annotators (A,B,C) using Cohen Kappa} 
\label{tab:cohenannotation2}
\vspace{-2em}
\end{wraptable}

\vspace{-0.5em}
\subsection{Annotation}
\vspace{-0.5em}
Capturing truly toxic content on social media for humans requires reliable annotation guidelines for training annotators. Our annotators have completed a rigorous training process including  literature reviews and discussions on online toxic behavior and its socio-cultural context among adolescents. Three annotators labeled the interactions using  three labels: \emph{Toxic (T), Non-Toxic (N)} and \emph{Unclear (U)}. The annotators were trained by our co-author cognitive scientist to consider the context of the interaction rather than individual tweets while determining the label of an interaction. We developed a guideline for annotators to follow that comprises intent-oriented criteria for labeling interactions as Toxic (T). That is, a tweet is toxic if the interactions contain: (i) Threat to harm a person, (ii) Effort to degrade or belittle a person, (iii) Express dislike towards a person or a group of people, (iv) Promote hate/violence/offensive language towards a person or a group of people, (v) Negatively stereotype a person or a minority, (vi) Support and defend xenophobia, sexism or racism.

\begin{table}[!htbp]
\vspace{-1em}
        \begin{minipage}{0.5\textwidth}
            \centering
            \begin{tabular}{p{1.8cm}p{1.cm}p{1.cm}p{1.cm}}
            \toprule[1.5pt]
             \textbf{Number of Tweets} & \textbf{Mean} & \textbf{Min} & \textbf{Max} \\ \midrule
              Toxic & 13.28 & 3.0 & 304.0 \\ \midrule
              Non-Toxic & 7.15 & 3.0 & 99.0 \\
             \bottomrule[1.5pt]
            \end{tabular}
            \subcaption{}
            \label{tab:tweetdatadistribution}
        \end{minipage}
        \hfill
        \begin{minipage}{0.5\textwidth}
            \centering
            \begin{tabular}{p{1.8cm}p{1.cm}p{1.cm}p{1.cm}}
            \toprule[1.5pt]
              \textbf{Number of Emoji} & \textbf{Mean} & \textbf{Min} & \textbf{Max} \\ \midrule
              Toxic & 6.72 & 0.0 & 290.0 \\ \midrule
              Non-Toxic & 3.51 & 0.0 & 60.0 \\
             \bottomrule[1.5pt]
            \end{tabular}
            \subcaption{}
            \label{tab:emojidatadistribution}
        \end{minipage}
        \vfill
        \begin{minipage}{0.5\textwidth}
            \centering
           \begin{tabular}{p{1.8cm}p{1.cm}p{1.cm}p{1.cm}}
            \toprule[1.5pt]
              \textbf{Number of URLs} & \textbf{Mean} & \textbf{Min} & \textbf{Max} \\ \midrule
              Toxic & 2.70 & 0.0 & 73.0 \\ \midrule
              Non-Toxic & 1.63 & 0.0 & 26.0 \\
             \bottomrule[1.5pt]
            \end{tabular}
            \subcaption{}
            \label{tab:urlsdatadistribution}
        \end{minipage}
        \hfill
        \begin{minipage}{0.5\textwidth}
            \centering
            \begin{tabular}{p{1.8cm}p{1.cm}p{1.cm}p{1.cm}}
                \toprule[1.5pt]
                  \textbf{Number of Images} & \textbf{Mean} & \textbf{Min} & \textbf{Max} \\ \midrule
                  Toxic & 1.18 & 0.0 & 20.0 \\ \midrule
                  Non-Toxic & 0.86 & 0.0 & 12.0 \\
                 \bottomrule[1.5pt]
            \end{tabular}
            \subcaption{}
            \label{tab:imagessdatadistribution}
        \end{minipage}
\vspace{-1em}
        \caption{(a)Descriptive statistics of tweets per interaction. (b) Descriptive statistics of emoji per interaction. (c) Descriptive statistics of URLs per interaction. (d) Descriptive statistics of images per interaction. There were 140 images showing Toxic Behavior and 471 images showing Non-Toxic Behavior.}
        \label{tab:tinv}
\vspace{-1.5em}
\vspace{-4mm}
\end{table}

If an annotator could not arrive at a conclusion after assessing the interaction following this guideline, it was labeled as \emph{Unclear}. After the annotations were completed by \emph{the three annotators}, the labels were finalized by majority vote. Then, agreement scores were computed utilizing Krippendorff’s alpha ($\alpha$) and Cohen’s Kappa ($\kappa$). Note that the instances labelled Unclear (U) can be included in the training to exercise the robustness of a learned model, or they can be removed as they add noise (as per the consensus of the annotators). To accommodate both scenarios, we create two schemes: (i) three label (T, N, U), (ii) two label (T, N) removing Unclear (U) instances \cite{gaur2019knowledge}. We perform two annotation analysis for both schemes: (i) A group-wise annotator agreement to find the robustness of the annotation by the three annotators using Krippendorff’s alpha ($\alpha$) \cite{soberon2013measuring}, (ii) A pair-wise annotator agreement using Cohen’s Kappa ($\kappa$) to identify the annotator with highest agreement with others. In the three-label scheme, $\alpha$ was computed as 0.63, and for the two label scheme, ($\alpha$) was 0.65.  The agreement scores reported in Table \ref{tab:kannotation} imply substantial agreement\footnote{\url{http://homepages.inf.ed.ac.uk/jeanc/maptask-coding-html/node23.html}} \cite{carletta1997reliability}. We also computed the agreement between annotators using $\kappa$ and provided in Table \ref{tab:cohenannotation1} and Table \ref{tab:cohenannotation2}, for three label and the two label, respectively. While the annotators A and B have substantial and near perfect agreement, C has moderate and substantial agreement with A and B, both for the three and two label schemes respectively \cite{carletta1997reliability}.

\begin{wraptable}{R}{0.51\textwidth}
\vspace{-2em}
\begin{tabular}{p{2.1cm}p{2.1cm}p{1.8cm}} \toprule[1.5pt]
     \textbf{Toxic} & \textbf{Non-Toxic} & \textbf{Unclear} \\
     \midrule
     118 (17.15\%) & 547 (79.51\%) & 23 (3.34\%)\\
     \bottomrule[1.5pt]
\end{tabular}
\vspace{-3mm}
\caption{\footnotesize Overall distribution of the data instances over the three labels.} 
\label{tab:datadistribution}

\vspace{1.5em}

\begin{tabular}{p{3cm}p{3cm}} \toprule[1.5pt]
    \textbf{Type of URLs} & \textbf{Number of URLs}  \\
    \midrule
    Image URLs & 140 (43.88\%) \\ 
    \midrule
    Video URLs & 44 (13.79\%) \\ 
    \midrule
    Text URLs & 48 (15.04\%) \\
    \bottomrule[1.5pt]
\end{tabular}
\vspace{-3mm}
\caption{\footnotesize Different types of URLs in \textbf{toxic} interactions.} 
\label{tab:typesofurlToxic}
\vspace{-1em}
\end{wraptable}

\subsection{Descriptive Statistics}
\vspace{-1mm}
In this section, we provide descriptive statistics of the dataset concerning the distribution of tweets, images, emoji and URLs with respect to labels. Table \ref{tab:datadistribution} shows the overall distribution of the instances as Toxic interactions constitute the 17.15\% of the dataset, while 79.51\% remains as Non-Toxic. 
A minority group of interactions with 3.34\% comprises the Unclear instances where annotators agreed that no conclusion could be derived. While the imbalance in the dataset  provides challenges in the modeling of toxic behavior, it is reflective of the nature of occurrence in real life. On the other hand, although the number of toxic interactions is smaller, they are richer in content as well as multimodal elements, compared to non-toxic interactions \cite{kursuncu2018s} (see Tables \ref{tab:tweetdatadistribution}, \ref{tab:emojidatadistribution}, \ref{tab:urlsdatadistribution}, \ref{tab:imagessdatadistribution}, and \ref{tab:typesofurlToxic}). Prior research shows that appropriate incorporation of multimodal elements in modeling with social media data would improve performance 
\cite{kursuncu2018s,kursuncu2019predictive,duong2017multimodal,o2014role}. 
In Table \ref{tab:tweetdatadistribution}, we see mean and maximum number of tweets per interaction for Toxic ones being significantly higher than Non-toxic ones, suggesting the intensity of the toxic content. Further, according to Tables \ref{tab:tweetdatadistribution}, \ref{tab:emojidatadistribution}, \ref{tab:urlsdatadistribution}, \ref{tab:imagessdatadistribution}, and \ref{tab:typesofurlToxic}, in the Toxic content, the use of multimodal elements such as image, video, and emoji, is clearly higher, suggesting that the incorporation of these different modalities in the analysis of this dataset will be critical for a reliable outcome \cite{kursuncu2019predictive,kursuncu2018s,duong2017multimodal,o2014role}.

\section{Discussion and Conclusion}
\label{sec:disc-conc}
\vspace{-2mm}
We created and examined the multimodal ALONE dataset for adolescent participants utilizing a lexicon \cite{rezvan2018quality} that divides offensive language into different types concerning appearance, intellectual, political, race, religion, and sexual preference. Given its unique characteristics concerning (i) adolescent population and (ii) interaction-based design, this dataset is an important contribution to the research community, as ground truth to provide a better understanding of online toxic behavior as well as training machine learning models \cite{salminen2018anatomy,kursuncu2019knowledge} and performing time-series analysis. Specifically, quantitative as well as qualitative analysis of this dataset will reveal patterns with respect to social, cultural and behavioral dimensions \cite{parent2019social,wandersman1998urban,safadi2020curtailing} and shed light on etiology of toxicity in relationships. Further, researchers can develop guidelines for different kinds of toxic behavior such as harassment and hate speech, and annotate the dataset accordingly. Lastly, we reiterate that the ALONE dataset will be available upon request to the authors, and the researchers will be required to sign an agreement to use it only for research purposes and without public dissemination.

\section*{Acknowledgement}
\vspace{-3mm}
We acknowledge partial support from the National Science Foundation (NSF) award CNS-1513721: “Context-Aware Harassment Detection on Social Media". Any opinions, conclusions or recommendations expressed in this material are those of the authors and do not necessarily reflect the views of the NSF.

\bibliographystyle{splncs04}
\vspace{-1mm}
\bibliography{reference.bib}
\end{document}